\documentclass{article}
\usepackage[utf8]{inputenc}
\usepackage{latexsym,amsmath,amssymb,amsfonts,amsthm,mathrsfs,bm,graphicx}
\usepackage{array}
\usepackage[table]{xcolor}
\usepackage{multicol}
\usepackage{multirow}
\usepackage{float,ccaption,caption,subcaption,subfloat} 

\usepackage{authblk}

\usepackage{soul}


\usepackage[margin={1.6cm,1.4cm}]{geometry}
\usepackage{color}
\usepackage{outlines}
\usepackage{enumitem}
\usepackage{soul}


\newcommand{\trb}[2]{\text{Tr}_{#1}\left[#2\right]}
\newcommand{\ket}[1]{| #1 \rangle}
\newcommand{\bra}[1]{\langle #1 |}

\newcommand{\ketbra}[1]{\ket{#1}\bra{#1}}

\newcommand{\identity}{\mathbb{I}}
\newcommand{\hilbert}{\mathcal{H}}

\newcommand{\jrf}[1]{\textcolor{green}{#1}}

\newcommand{\yudong}[1]{\textcolor{red}{#1}}

\setenumerate[1]{label=\Roman*.}
\setenumerate[2]{label=\Alph*.}
\setenumerate[3]{label=\roman*.}
\setenumerate[4]{label=\alph*.}

\title{QVECTOR: an algorithm for device-tailored quantum error correction}
\author[1]{Peter D. Johnson}
\author[1]{Jonathan Romero}
\author[1]{Jonathan Olson}
\author[1]{Yudong Cao}
\author[1,2]{Al\'{a}n Aspuru-Guzik}
\affil[1]{Department of Chemistry and Chemical Biology, Harvard University, 12 Oxford Street, \protect\\ Cambridge, MA 02138, USA}
\affil[2]{Senior Fellow, Canadian Institute for Advanced Research, Toronto, Ontario M5G 1Z8, Canada}

\begin{document}
\maketitle

\noindent \textbf{Current approaches to fault-tolerant quantum computation will not enable useful quantum computation on near term devices of 50 to 100 qubits. Leading proposals, such as the color code and surface code schemes, must devote a large fraction of their physical quantum bits to quantum error correction. Building from recent quantum machine learning techniques, we propose an alternative approach to quantum error correction aimed at reducing this overhead, which can be implemented in existing quantum hardware and on a myriad of quantum computing architectures. 
This method aims to optimize the average fidelity of encoding and recovery circuits with respect to the \emph{actual} noise in the device, as opposed to that of an artificial or approximate noise model. 
The quantum variational error corrector (QVECTOR) algorithm employs a quantum circuit with parameters that are variationally-optimized according to processed data originating from quantum sampling of the device, so as to learn encoding and error-recovery gate sequences. We develop this approach for the task of preserving quantum memory and analyze its performance with simulations. We find that, subject to phase damping noise, the simulated QVECTOR algorithm learns a three-qubit encoding and recovery which extend the effective $T_2$ of a quantum memory six-fold.
Subject to a continuous-time amplitude- plus phase-damping noise model on five qubits, the simulated QVECTOR algorithm learns encoding and decoding circuits which exploit the coherence among Pauli errors in the noise model to outperform the five-qubit stabilizer code and any other scheme that does not leverage such coherence. 
Both of these schemes can be implemented with existing hardware.
}

\begin{multicols}{2}

Uncontrollable environmental errors have remained the primary roadblock on the route to useful quantum information processing. 
Nevertheless, there is hope for achieving fault-tolerant quantum computation by implementing quantum error correction \cite{Calderbank1996,Knill1997,Gottesman1997,Dennis2002}. Fault-tolerant threshold theorems \cite{Aharonov2008} show that, for a given degree of environmental noise, if each elementary operation can perform below a certain error rate, then concatenated quantum error correction schemes will out-pace error accumulation, enabling quantum computation to an arbitrary degree of accuracy.

The leading approaches to quantum error correction make use of topological stabilizer codes \cite{Dennis2002,Bombin2006}. A major advantage of this approach is that the measurements used to diagnose errors may be performed on just a few neighboring qubits. 
Leading candidates for topological quantum error correction are the surface code \cite{Dennis2002,Fowler2012}, color code \cite{Bombin2006}, and gauge color code \cite{Bombin2015}. Progress towards implementing surface and color codes experimentally has been demonstrated in \cite{Barends2014,Hill2015} and \cite{Nigg2014}, respectively. Recent simulations have shown that the gauge color code \cite{Brown2016} is expected to exhibit performance comparable to the previous schemes, though with the added benefit of transversal implementation of a universal gate set.

Unfortunately, these codes are not likely to be practical in near-term devices. Current estimates predict that the surface code will require $10^3$ to $10^4$ physical qubits to protect a single logical qubit \cite{Fowler2012}. In order to perform useful, fault tolerant quantum computation with near term devices, it seems that this number of physical qubits needs to be drastically reduced. We propose a path towards reducing such error-correction overhead via the design of device-tailored quantum error correcting codes.


In an actual device, quantum information is subject to hardware-specific quantum noise processes. Stabilizer codes are not optimal, in general, because they are not tailored to the specific noise of a given device \cite{Leung1997}. Significant work in quantum error correction has investigated codes outside of the stabilizer formalism, which are tailored to noise beyond the Pauli error model \cite{Chuang1996,Cirac1996,Leung1997,Reimpell2005,Cafaro2014}. Various schemes have been developed to numerically optimize encoding and decoding procedures with respect to a fixed noise model \cite{Fletcher2008IEEE,Fletcher2008PRA,Kosut2008,Kosut2009,Taghavi2010}. 

However, these optimization schemes are not applicable to useful quantum processing devices. First, they require a specific noise model as input to the optimization. Significant effort \emph{has} been devoted to characterizing the noise of near-term devices \cite{Koch2007,Bylander2011,Peterer2015} and quantum error correcting codes have been implemented as a tool for such characterization \cite{Laforest2007}. But, as larger systems are considered, the accuracy of these noise models is expected to drop rapidly \cite{Katabarwa2017}.

Second, even if a sufficiently accurate noise model were known, existing classical processors are unable to handle the storage needed to carry out such optimization for near-term devices with 50 qubits \cite{Boixo2016}. The task of performing such an optimization seems better suited for a quantum processor.

Finally, even in cases where these optimizations can be performed, the optimized encoding and recovery processes must be decomposed into a sequence of gates that are available on the device. It may be favorable, rather, to employ an optimization strategy which naturally integrates the constraints of the device's native gate set.

Recently, several hybrid quantum-classical (HQC) algorithms for solving specific optimization tasks have been developed. Two representative variational HQC algorithms are the variational quantum eigensolver (VQE) \cite{Peruzzo2014} and the quantum approximate optimization algorithm (QAOA) \cite{Farhi2014}. The former of these algorithms has been implemented experimentally on several quantum computing architectures \cite{Peruzzo2014,Shen2017,OMalley2016, Kandala17}. Additionally, much theoretical work has been done to develop this genre of quantum algorithms \cite{Wecker2015,McClean2016,Wang2017,Guerreschi2017,McClean2017,Romero2017}.
A major appeal to such algorithms is that they operate successfully without the need for active quantum error correction and even show signs of suppressing certain types of errors \cite{McClean2016,McClean2017}.

Variational HQC algorithms are implemented by preparing quantum states as the output of a parameterized quantum circuit $U(\vec{p})$. Various ansatz states are repeatedly prepared and measured to collect outcome samples. The measurement data is classically processed and used to update the circuit parameters so as to optimize a particular cost function. As in the quantum autoencoder algorithm \cite{Romero2016}, the variational optimization of a circuit constitutes a quantum analogue of training a neural network in machine learning.

We propose a variational HQC algorithm for designing device-tailored error corrected quantum memories. This approach naturally points to an extension for designing error corrected gates. The algorithm solves a number of problems which hamper the classical optimization schemes for the same task. First and foremost, our proposal forgoes the need for a noise model because the optimization is carried out \emph{in situ} and the noise perfectly simulates itself on the device. Second, the optimization step is not necessarily hindered by the exponential scaling of the Hilbert space dimension in the same way that the previous proposals are. Measurement statistics, obtained using the device as a quantum sampler, are used to approximate the average fidelity of the encoding-decoding scheme. This average fidelity serves as the cost function for the classical optimization algorithm. Finally, by constructing the variational circuits out of a device-native gate set, the optimized encoding and recovery processes are manifestly decomposed into an implementable sequence of gates.

\vspace{.5cm}

\noindent\textbf{Quantum error correction}

 \noindent The Pauli group stabilizer formalism of quantum error correction \cite{Gottesman1997} has earned its place as the most popular approach to quantum error correction. By describing all mathematical objects in terms of elementary gates, the formalism has enabled significant theoretical analysis and experimental implementation. Our proposal, however, is not based on the Pauli group stabilizer formalism. Accordingly, we review a more general framework of quantum error correction. 

The general mathematical formalism of subspace code quantum error correction \cite{Knill2006} is summarized as follows. First, $k$ qubits of logical quantum information $\hilbert_L\simeq \mathcal{Q}^{\otimes k}$ are encoded via an encoding process $\mathcal{E}$ into $n$ physical qubits $\hilbert_P\simeq \mathcal{Q}^{\otimes n}$. Next, the physical qubits are subjected to some noise process $\mathcal{N}$. Finally, the quantum information is attempted to be recovered by a decoding process $\mathcal{D}$. The quality of the quantum error correction scheme can be characterized by how well the sequence of processes approximates the identity channel $\mathcal{D}\circ\mathcal{N}\circ\mathcal{E}\approx \mathcal{I}$, which may be quantified by either the average fidelity or worst-case fidelity of the quantum process, or by some other figure of merit.

It is standard to use an encoding in which encoded states are pure: $\mathcal{E}(\ketbra{\psi})=\ketbra{\overline{\psi}}$. 
The linear span of state vectors in the range of $\mathcal{E}$ is referred to as the code space $\mathcal{C}$. The code space is a $2^k$-dimensional subspace of the physical Hilbert space, $\mathcal{C}\leq\hilbert_P$. It is instructive to factor the code space into a logical subsystem $\hilbert_{\bar{L}}$ and a syndrome subsystem $\hilbert_S$,
\begin{equation}
\mathcal{C}\simeq \hilbert_{\bar{L}}\otimes\textup{span}(\ket{s_0})\leq\hilbert_P\simeq\hilbert_{\bar{L}}\otimes\hilbert_S,
\end{equation}
where $\ket{s_0}$ is a fixed state of $\hilbert_S$. This factorization does not correspond to a separation of physical qubits, but rather to a separation of \emph{virtual subsystems} \cite{Zanardi2001, Poulin2005}, and is the key structure of the \emph{subsystem principle} of quantum error correction \cite{Knill2006}.

The encoding should be chosen to match the features of the noise model. It is standard to model the noise as a completely positive trace-preserving map, expressed in Kraus representation as $\mathcal{N}(\cdot)=\sum_j K_j\cdot K_j^{\dagger}$. Perfect recovery is possible if and only if there exists an encoding $\mathcal{C}$ such that each Kraus operator satisfies
\begin{equation}
K_j\ket{\overline{\psi}}=V^{\dagger}(\ket{\psi}\otimes\ket{\tau_j})
\end{equation}
for all $\ket{\overline{\psi}}\in\mathcal{C}$ and for a fixed unitary $V^{\dagger}$ and some unnormalized $\ket{\tau_j}$ which depend on $K_j$; note that this is simply a rephrasing of the Knill-Laflamme condition \cite{Knill1997} for exact correctability. Conditional upon the syndrome system $\hilbert_S$ being initialized in $\ket{s_0}$, the logical quantum information is protected from the noise in the virtual subsystem $\hilbert_{\bar{L}}$. If errors are to be corrected while the quantum information is still encoded, $V^{\dagger}$ is inverted by the application of $V$ and the syndrome system is reset back to $\ket{s_0}$. To decode, the logical subsystem is mapped back into the $k$-qubit system $\hilbert_L$, and the syndrome qubits are simply traced out.
 
To briefly make contact with the stabilizer formalism, the code is defined by the intersection of eigenvalue-1 subspaces of the Pauli stabilizer operators $\{S_j\}$, which admit the decomposition $S_j=\identity_{\bar{L}}\otimes (\tilde{S}_j)_S$, while the logical operators $\{\overline{Z}_i,\overline{X}_i\}$ of the code decompose as $\overline{Z}_i=(Z_i)_{\bar{L}}\otimes\identity_S$. For a more thorough account of this connection, see \cite{Poulin2005} or \cite{Knill2006}.

\vspace{.5cm}

\noindent \textbf{Quantum variational error correction algorithm}

\begin{figure*}[ht]
\centering
\includegraphics[scale=0.27]{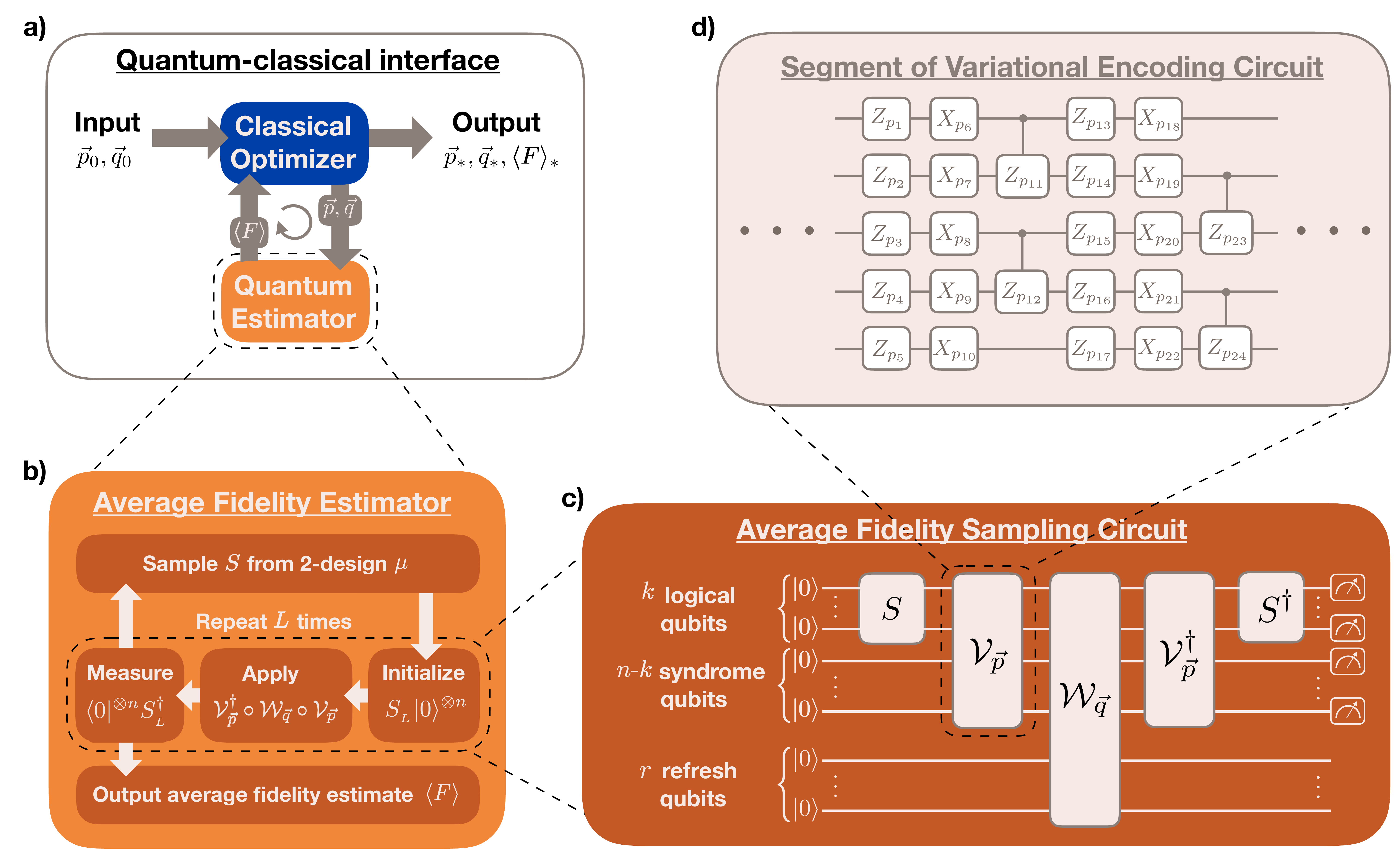}
\vspace*{0cm}
\caption{Training stage of the QVECTOR algorithm. a) QVECTOR uses a classical optimizer to optimize a function whose output value is determined by calling a quantum subroutine, the quantum average fidelity estimator. b) The quantum average fidelity estimator 
uses a variational quantum circuit to send random-sampled states $S\ket{0}^{\otimes k}$ through the circuit $\mathcal{W}_{\vec{q}}\circ\mathcal{V}_{\vec{p}}$ and records the measured channel fidelity of each state as $0$ or $1$. The average of these measured channel fidelities is output and fed back to the classical optimizer. c) Within each call to the average fidelity estimator, the quantum circuit is run $L$ times. In each run, the state $\ket{0}^{\otimes n}$ is initialized on $k$ logical qubits and $n-k$ syndrome qubits, the $k$ logical qubits are transformed by the 2-design sampled unitary $S$, the noisy encoding-decoding circuits $\mathcal{V}_{\vec{p}}$ and $\mathcal{W}_{\vec{q}}$ are applied, the inverse of $S$ is applied, and the $k$ logical qubits are measured in the computational basis.
d) The variational circuit $\mathcal{V}_{\vec{p}}$ and $\mathcal{W}_{\vec{q}}$ may be parameterized as seen fit by the particular device. One example of $\mathcal{V}_{\vec{p}}$, constructed from single-qubit X- and Z-rotations, and 2-qubit controlled-Z rotations is depicted; here, the variational parameters, $p_1,p_2,\ldots$, are the rotation angles of each circuit element.
}
\label{fig:qvschematic} 
\end{figure*}

\noindent The objective of the quantum variational error corrector (QVECTOR) algorithm is to output an encoding and recovery circuit which sufficiently correct errors in a quantum memory. First we describe the protocol used for optimizing, or \emph{training}, the encoding-recovery circuits, then we will describe their implementation as an error correction scheme.

The physical qubits are divided into two sets, qubits which will encode $k$ logical qubits and $r$ qubits which facilitate the non-unitary error-recovery process. The encoding and recovery are implemented with a circuit of tunable gates. 
Before encoding, the first $k$ qubits are prepared in a to-be-encoded state $\ket{\psi}$ and the remaining $n-k$ qubits are initialized in a fiducial state $\ket{0}^{\otimes n-k}$. The sequence of gates $V(\vec{p})$ then acts to encode the state of the first $k$ qubits into $n$ qubits.
The recovery process is aided by an additional register of $r$ ``refresh'' qubits, and is implemented by a sequence of gates $W(\vec{q})$
coupling all $n+r$ qubits. 



The figure of merit, or cost function, we use to evaluate an encoding-recovery pair is the \emph{average code fidelity}, based on the quantum channel Haar average fidelity \cite{Dankert2009}. 
 For a (possibly) noisy recovery operation $\mathcal{R}$, the average code fidelity is defined as
\begin{align}
\label{eq:avgfid}
\langle F\rangle &\equiv \int_{\psi\in\mathcal{C}}\bra{\psi}\mathcal{R}\left(\ketbra{\psi}\right)\ket{\psi}d\psi,
\end{align}
where the integral is performed with respect to the Haar distribution of states in the code space.
In addition to evaluating the preservation of the logical qubits, it scores the ability of the encoded recovery circuit $W(\vec{q})$ to properly return the quantum information to the code space. A well-performing encoded recovery operation can be applied in sequence to extend the lifetime of the quantum memory.

As a hybrid quantum-classical algorithm, QVECTOR uses a quantum and a classical processing unit working in tandem.
The objective of the classical processing unit is to \emph{optimize} the average code fidelity with respect to the tunable circuit parameters $(\vec{p},\vec{q})$, while the quantum information processing unit is called by the classical processor as a subroutine to \emph{estimate} the average code fidelity associated with the given encoding-recovery pair $(V(\vec{p}),W(\vec{q}))$.  

The average code fidelity estimation procedure we use is inspired by the the sampling approach of \cite{Emerson2005,Dankert2009}. 
Assuming $S$ and $S^{\dagger}$ are efficient to implement, the input-output fidelity of any term in Eq.~(\ref{eq:avgfid}) can be efficiently computed by preparing $\ket{0}^{\otimes k}\ket{0}^{\otimes n-k}$, applying state preparation $S$ on the first $k$ qubits, performing the encoding-decoding operation $(V(\vec{p})^{\dagger}_n\otimes\identity_r)W(\vec{q})(V(\vec{p})_n\otimes\identity_r)$, applying the inverse of $S$, and then measuring the first $n$ qubits in the computational basis. After $N$ trials, the fraction of all-$0$ outcomes estimates the fidelity of $\mathcal{R}(\ketbra{\psi})$ with respect to $\ketbra{\psi}$ with standard deviation $\mathcal{O}(\frac{1}{\sqrt{N}})$. 

To obtain an estimate of the \emph{average} code fidelity, we could vary the state preparation circuit $S$, and obtain code fidelity data for sufficiently many samples $S$ drawn from the Haar distribution. However, because Haar-random sampling is not efficient \cite{Nielsen2002} and because the average code fidelity depends only on the second moment of the distribution, we instead sample $S$ from an efficiently implementable unitary 2-design \cite{Dankert2009}.
A unitary $2$-design is a measure $\mu$ on the unitary group $\mathcal{U}(d)$ satisfying 
\begin{equation}
\int_{\mathcal{U}(d)}S^{\otimes 2}\otimes {S^{\dagger}}^{\otimes 2}d\mu(S)=\int_{\mathcal{U}(d)} U^{\otimes 2}\otimes {U^{\dagger}}^{\otimes 2}dU.
\end{equation}
With a $2$-design $\mu$, the average code fidelity of the encoding-decoding is written as
\begin{align}
\label{eq:avgcodefid}
\langle F\rangle &\equiv \int_{\mathcal{U}(d)}\bra{0^{(n)}}S^{\dagger}\mathcal{V}_{\vec{p}}^{\dagger}\mathcal{W}_{\vec{q}}\mathcal{V}_{\vec{p}}\left(S\ketbra{0^{(n)}}S^{\dagger}\right)S\ket{0^{(n)}}d\mu(S),
\end{align}
where $\mathcal{V}_{\vec{p}}$ and $\mathcal{W}_{\vec{q}}$ denote the physically implemented quantum channels---noisy versions of the parameterized circuits. This quantity may be estimated with standard deviation $N$ after $\mathcal{O}(\frac{1}{\sqrt{N}})$ trials as follows. In each trial, $S$ is sampled from the 2-design and implemented in the process $\mathcal{S}^{\dagger}\mathcal{V}_{\vec{p}}^{\dagger}\mathcal{W}_{\vec{q}}\mathcal{V}_{\vec{p}}\mathcal{S}$ that is applied to the initial state $\ket{0}^{\otimes n}$. In each trial, the first $n$ qubits are measured in the computational basis and the number of all-0 outcomes over $N$ constitutes an unbiased estimator for the average code fidelity.


In some cases, it may be favorable to implement an approximate unitary 2-design. A good candidate is the recent construction of an $\epsilon$-approximate 2-design \cite{Nakata2017}, which is particularly simple to implement. Using this approximate 2-design, each quantum measurement is interpreted as returning a binary sample from a \emph{biased estimator} of the true average fidelity. As shown in Appendix \ref{app:fidelityest}, using $\ell$ applications of the randomization circuit in \cite{Nakata2017}, the bias of this estimator is upper bounded by $\frac{2^{k(\ell+1)}+2^{k\ell}-2}{2^{2k\ell}(2^{k}-1)}\sim \mathcal{O}(1/2^{k\ell})$. Thus, after $N$ samples from this biased estimator, the estimated average fidelity is expected to deviate from the true average fidelity as $\mathcal{O}(\frac{1}{\sqrt{N}}+\frac{1}{2^{k\ell}})$.

A schematic of this fidelity estimation algorithm is depicted in Figure \ref{fig:qvschematic}a. The quantum fidelity estimation algorithm serves as a cost function evaluation subroutine which is called by the classical processor that performs an optimization such as LBFGS \cite{Wright1999} or SPSA \cite{Spall1992}. Any ``quantum speedup'' realized by this algorithm is due to the efficiency with which a quantum circuit can be used to estimate its own average fidelity\footnote{Note that the analysis in this paper does not unequivocally prove that such a speedup is possible with this approach. As this method is not easily amenable to theoretical analysis, a proper evaluation of its effectiveness will come from physical implementation.}.

After the circuit is trained to a sufficient average fidelity, it is ready to be used as a quantum error correction scheme for preserving a quantum memory. Once the quantum information is encoded using $V(\vec{p})$, the recovery circuit $W(\vec{q})$ may be applied at regular time intervals to recover from errors accrued in the memory. Before each recovery step, the refresh qubits must be reset to the fiducial state $\ket{0}^{\otimes r}$. 

\vspace{.5cm}

\noindent \textbf{Simulation results}
\begin{figure*}[ht!]
\centering
\begin{minipage}[t]{.48\textwidth}
        \centering
        \includegraphics[width=\linewidth]{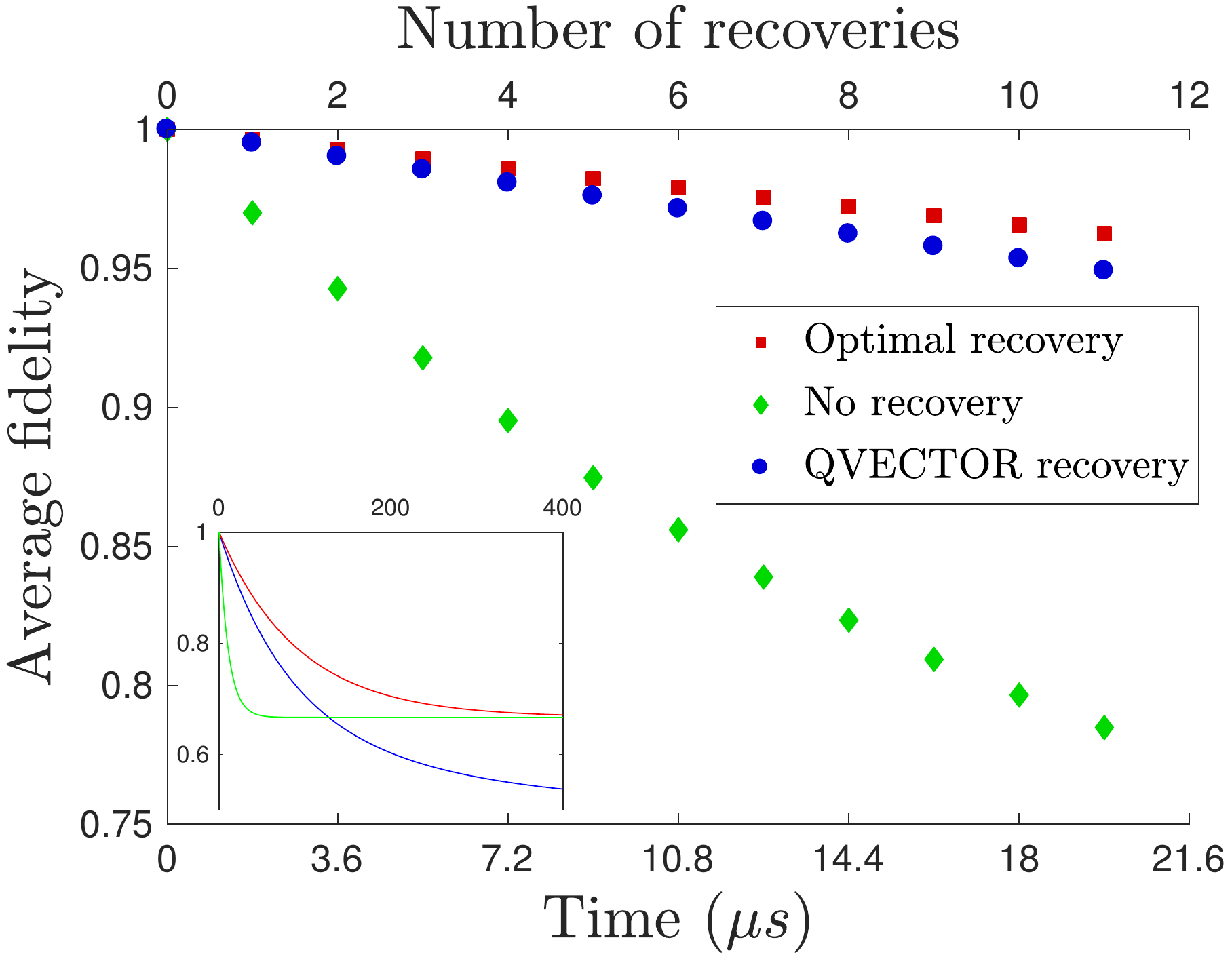}
        \vspace*{-0.5cm}
        \caption{The average fidelities after each 1.8 $\mu s$ recovery step are plotted for various procedures, showing that the QVECTOR-learned recovery nearly matches that of the optimal recovery procedure, as given by the standard phase damping code. In the no-recovery case (i.e. decoherence of a single physical qubit), the noise process in each step is modeled by a probabilistic phase damping with error probability $p=0.045$, corresponding to $T_2= 19~\mu s$ \cite{Barends2014}. To account for additional error due to noisy gates in the optimal and QVECTOR recoveries, the noise here is modeled by a probabilistic phase damping with error probability $p=0.091$. Despite the addition of gate error, the QVECTOR recovery extends the effective $T_2$ by nearly six-fold to $\sim 110~\mu s$, while the optimal recovery extends this to $\sim 165~\mu s$. The QVECTOR recovery circuit uses just 30 layers of two-qubit gates, which is comparable to the number used in the optimal recovery circuit. The inset depicts the many-recovery limit, where the QVECTOR average fidelity eventually drops below the no-recovery average fidelity after roughly 150 recovery steps, possibly due to systematic over-rotation in the learned recovery process.} 
\label{fig:recoveries}
   \end{minipage}
    \hspace*{.02\textwidth}
    \begin{minipage}[t]{.48\textwidth}
        \centering
        \includegraphics[width=\linewidth]{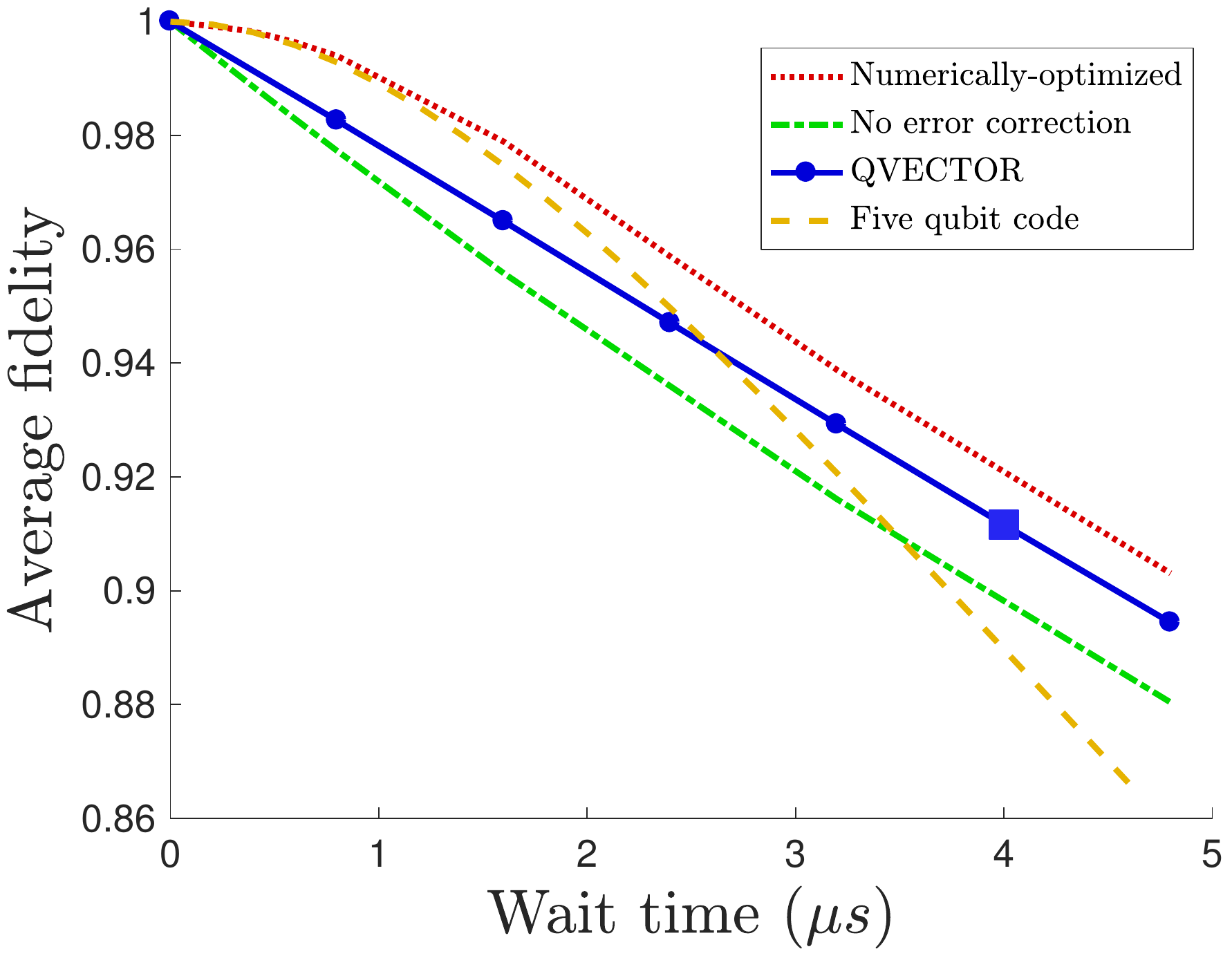} 
        \vspace*{-0.5cm}
        \caption{We consider a quantum communication setting, where only a single recovery-decoding step is available after a pre-determined ``wait time''. For each wait time, the average fidelities of various encoding-decoding procedures are plotted, showing that the QVECTOR-learned encoding-decoding scheme continues to be useful beyond the time at which the standard five-qubit code fails to be so. Between encoding and decoding, the noise is modeled as a continuous amplitude- plus phase-damping channel with $T_2=19~\mu s$ and $T_1=57~\mu s$ (see Appendix \ref{app:noisesimulation}). The numerically-optimized encoding-decoding pairs are obtained using the iterated semi-definite programming method of \cite{Kosut2009}. The QVECTOR-learned encoding-decoding pair were initially trained for the $4~\mu s$ wait time. This QVECTOR-optimized encoding-decoding pair for $4~\mu s$ was used as an initial point for gradient-based optimization of the remaining wait times. The QVECTOR encoding-decoding pairs continue to be useful beyond the point at $3.5~\mu s$ where the five-qubit code drops below the no-encoding average fidelity.} \label{fig:gellerplot}
    \end{minipage}
\end{figure*}

Towards evaluating the effectiveness of this algorithm, we simulate its performance on several few-qubit examples in the presence of simple noise models. Against three qubit phase-damping error, we find that the algorithm is able to learn an encoding and recovery map which perform nearly as well as the optimal phase-error code and recovery process. 
Considering a more realistic noise process by incorporating amplitude damping \cite{Geller2013} on five qubits, we find that our algorithm can learn useful error correction which exploits coherence in the Pauli errors where the five-qubit code fails to improve the physical-qubit fidelity.

All noise in the system is modeled as a quantum channel which acts after the encoding map and before the recovery process. As such, the state preparation, parameterized circuit gates, and measurements are taken to be ideal. For a noise channel $\mathcal{N}$, the optimization cost function is defined as the average code fidelity of the quantum process $\mathcal{V}_{\vec{p}}^{\dagger}\mathcal{W}_{\vec{q}}\mathcal{N}\mathcal{V}_{\vec{p}}$.

The variational circuits consist of layers of single qubit rotations interleaved with two-qubit entangling operations. An example is shown in Figure \ref{fig:qvschematic}.d., where the variational circuits consist of single-qubit Pauli-X and Pauli-Z rotations as well as nearest-neighbor controlled-Z rotations, whereby the variational parameters $\vec{p}$ and $\vec{q}$ control the rotation angles of each of these gates. The specific form of the variational circuits we use is described in Appendix~\ref{app:simulation}. As the QVECTOR algorithm is agnostic with respect to the choice of variational circuit structure, we chose to simulate circuits which can be implemented natively in existing hardware \cite{Barends2014}. 
The classical optimization can be performed using a variety of methods, including SPSA \cite{Spall1992}, basin-hopping \cite{Wales1997}, or Bayesian optimization \cite{Mockus2012}. However, the reported data was obtained using the quasi-Newton method L-BFGS \cite{Wright1999}. 

\textit{Three-qubit phase-damping-} In the first simulation, our goal is to analyze the performance of a quantum memory with active recovery learned by the QVECTOR algorithm. We consider a setting where a single qubit is encoded into three qubits which are subject to independent probabilistic phase damping. Two additional (noise-free) qubits are used to facilitate an encoded recovery operation. As described in Appendix \ref{app:noisesimulation}, the phase-flip error rate is chosen to match the specifications of a sequence of realistic one- and two-qubit gates implemented with Xmon qubits \cite{Barends2014} corresponding to the particular parameterized circuit we employ (see Appendix \ref{app:simulation}). We find that this corresponds to a single qubit phase-flip probability of $p=0.091$ and requires a duration of $1.8~\mu s$. There are two points of reference for assessing the performance of QVECTOR. The first is the case of no error correction, where a single physical qubit is used as a quantum memory. As outlined in Appendix \ref{app:noisesimulation}, to account for the lack of noisy gates in this case, the error rate in the no-encoding case is decreased to $p=0.045$ per time step. The second point of reference is the case of optimal encoding and recovery with respect to the $p=0.091$ phase-damping process. The standard phase-error code and corresponding recovery optimize the average fidelity metric with respect to a general phase-damping process.

As described in detail in Appendix \ref{app:simulation}, $V(\vec{p})$ and $W(\vec{q})$ are trained with respect to the above noise model. The encoding and recovery pair we use was selected as the optimally performing scheme among twelve distinct training attempts. To simulate the performance of these optimized circuits as a quantum memory, we compute the average fidelity of the process $\mathcal{V}^{\dagger}_{\vec{p}}(\mathcal{W}_{\vec{q}}\mathcal{N})^M\mathcal{V}_{\vec{p}}$ for a various number of iterations $M$. 
As shown in Fig. \ref{fig:recoveries}, we find that, with respect to phase-damping noise, the simulated QVECTOR algorithm results in a quantum memory that has an effective $T_2$ time of approximately $110~\mu s$, nearly six times that of the bare physical qubit. This shows that, although the gates used to implement the QVECTOR recovery circuit are modeled as to incur additional noise, there is, nonetheless, an improvement in the lifetime of the quantum memory. 

\begin{table*}[!]
\centering
\Large{Exploiting coherence between Pauli errors}\\

\vspace{0.2cm}

\large
\begin{tabular}{|c|c|c|}
\hline
\textbf{Scheme} & \textbf{Exact APD} & \textbf{PTA-APD} \\ 
\hline
$\textup{QVECTOR}^*$               &  \cellcolor{black!60!green} \textcolor{white}{0.912}     & 0.832   \\ 
\hline
Numerically optimized & 0.920     &  \cellcolor{yellow!40!red}\textcolor{white}{0.904}  \\ 
\hline
No-encoding           & \multicolumn{2}{c|}{0.898}        \\ 
\hline
	Five qubit code          &  \multicolumn{2}{c|}{\cellcolor{blue!15!red!80!black}\textcolor{white}{0.890}}   \\ 
\hline
\multicolumn{3}{l}{$^*$ \footnotesize The same encoding-decoding is used for both Exact APD and PTA-APD.}
\end{tabular}
\caption{This table compares the average fidelities achieved by various error correction schemes for the amplitude- plus phase-damping noise model at wait time $4~\mu s$. The right-most column reports the performance of the scheme against an approximate noise model, which ignores coherence among Pauli errors, obtained by Pauli-twirling \cite{Geller2013} the noise channel (see Table \ref{KrausOps}). While the five qubit code (red) clearly does not take advantage of the coherence in the amplitude- plus phase-damping (APD) channel, the encoding-decoding obtained by QVECTOR can be shown to genuinely make use of this coherence. The average fidelity achieved by the QVECTOR encoding-decoding (green) is, not surprisingly, diminished when computed against a Pauli-twirling approximation (PTA) of the APD noise. To show that the QVECTOR encoding-decoding unequivocally exploits the Pauli error coherence, we compare to the average fidelity of the numerically-optimized encoding-decoding, subject to the PTA-APD noise model, where Pauli-error coherence is removed (orange). The discrepancy between these two shows that the QVECTOR encoding-decoding exploits the coherence between Pauli errors in the noise model to achieve an average fidelity that could not be reached without such coherence.
}\label{QVECTORPTA}
\end{table*}

For the first ten or so recovery steps, the QVECTOR average fidelity remains comparable to that of the optimal recovery. However, as shown in the inset of Figure \ref{fig:recoveries}, the QVECTOR average fidelity equilibrates to $\langle F\rangle=1/2$ as opposed to $\langle F\rangle=2/3$, dropping below the no-encoding curve after roughly 150 iterations. We conjecture that this is due to a systematic over-rotation in the recovery process which accrues over many repeated recoveries. We obtained evidence for this explanation by examining the zero-noise limit and finding that the average fidelity in this case undergoes damped harmonic oscillation with a period of 1089 recovery steps. By training on just a single recovery step, such an over-rotation is indistinguishable from incoherent error. This points to the possibility of mitigating such over-rotation error by training on multiple recoveries, as discussed in the outlook section.

\textit{Five-qubit communication setting with amplitude- plus phase-damping error}- The goal of the second simulation is to test the simulated QVECTOR performance against a more realistic noise model and in a different error-correction setting. In some instances of a quantum memory, such as during transmission of a quantum state during communication, active error recovery is impractical or unavailable. This situation arises, for instance, if one were attempting to relay qubits through an optical fiber or transport qubits between two neighboring quantum processors. If one cannot repeatedly apply a recovery channel during transmission, the best error-reduction one can hope to achieve is an optimized encoding at the source followed by a single decoding at the destination.  We investigate the performance of the QVECTOR algorithm in such a case, determining the average fidelity for various ``wait times'' corresponding to the delay between transmission and reception when the state is subject to error.

In the single decoding scenario, the quantum information does not need to be returned to the code by the recovery step. Rather, the encoded quantum information only needs to be decoded back to the first physical qubit. Thus, a unitary correction suffices, and the refresh qubits are unnecessary (i.e. $r=0$). In this setting, we analyzed QVECTOR's performance where $k=1$, $n=5$, and $r=0$ subject to independent continuous amplitude- plus phase-damping (APD) for various wait times with $T_1=57~\mu s$ and $T_2=19~\mu s$, as described in Appendix~\ref{app:noisesimulation}. 

As shown in Fig.~\ref{fig:gellerplot}, the simulated QVECTOR algorithm learns an encoding and recovery pair with average fidelity greater than a physical qubit subject to the same noise process. We also compare to the standard five qubit code, which is known to be optimal for depolarizing noise \cite{Reimpell2005}. We find that, although the five qubit code fails to be useful after $t=3.5~\mu s$, by training QVECTOR for a 4$~\mu$s wait time, the encoding-decoding circuit learned by QVECTOR outperforms the no-encoding average fidelity for all wait times considered. The numerically-optimized average fidelity is obtained using the iterated semi-definite programming method of \cite{Kosut2009} (see Appendix \ref{app:simulation}), and plotted for comparison. Through training, the encoding-decoding pair was selected as the optimally performing scheme among three distinct training attempts.

Finally, we investigated the potential of the QVECTOR algorithm for discovering encoding-decoding circuits which exploit structure in the noise in ways that stabilizer codes do not. In the amplitude- plus phase-damping model, the Kraus operators are coherent superpositions of Pauli operators. A common technique in simulating noise is to ignore such coherences, and represent the process as an incoherent mixture of Pauli errors. This simplification is referred to as the Pauli-twirling approximation (PTA) \cite{Geller2013}. 

There is mounting evidence in the literature that, in order to compute the performance of stabilizer codes under realistic noise processes, it suffices to compute their performance against a PTA version of the corresponding noise channel \cite{Tomita2014,Gutierrez2015}. However, this condition holds only for small error rates \cite{Katabarwa2015}. 
At higher error rates and in the presence of coherent errors, error correction schemes constructed around the Pauli-error model can lead to fidelities even worse than those obtained without encoding (c.f. Fig.~\ref{fig:gellerplot}).

We considered the APD noise model acting for $4~\mu s$ with $T_1=57~\mu s$ and $T_2=19~\mu s$, matching the parameters used in Fig.~\ref{fig:gellerplot}. As found in Table \ref{KrausOps}, the Kraus operators of the APD channel are coherent superpositions of Pauli errors, while the Pauli-twirled approximation of the APD channel (PTA-APD) constitutes an incoherent mixture of Pauli errors. An important physical difference between these two channels is that the former is non-unital, enabling $T_1$ decay to the ground state.

As shown in Table \ref{QVECTORPTA}, we computed the average fidelity for various encoding-decoding schemes subject to APD and to PTA-APD. For the QVECTOR case we used the encoding-decoding obtained for the $4~\mu s$ APD noise model as reported in Fig. \ref{fig:gellerplot}. The performance of this scheme (green) relies significantly on the coherence between the Pauli errors in the APD noise model, as evidenced by the discrepancy between its APD and PTA-APD average fidelities. In contrast, the five qubit code does not utilize the coherence among the Pauli errors (red). It remains to verify that the average fidelity achieved by QVECTOR cannot be obtained by any encoding-decoding if the Pauli-error coherence is removed from the noise model. We performed the bi-convex optimization method of \cite{Kosut2009} to numerically determine the maximal average fidelity that can be achieved among all encoding-decoding schemes subject to PTA-APD (orange). Comparing this to the QVECTOR value, we find that the QVECTOR encoding-decoding exploits the coherence between Pauli errors in that this average fidelity cannot be achieved by any encoding-decoding scheme if Pauli-error coherence were removed from the noise process.


This finding highlights the fact that codes designed to be agnostic to coherence among Pauli errors, such as many stabilizer codes, fail to exploit such structures. 
In contrast, the QVECTOR methodology does not build in this limitation, as it does not assume a noise model \emph{a priori}. Therefore, this approach may be able to outperform other approaches by exploiting structure, such as Pauli-error coherence, in the noise processes.

\vspace{.5cm}

\noindent \textbf{Outlook}

We developed a hybrid quantum-classical algorithm which learns encoding and error-recovery processes tailored to the noise of the target quantum device. The opportunities for improvement over leading quantum error correction techniques are three-fold. First, by 
using a native parameterized gate set, this approach may facilitate a more-effective use of available resources for realizing quantum error correction.
Second, compared to previous optimization-based approaches, the optimization algorithm in QVECTOR is, in principle, scalable: the simulation of the noise process is efficient and accurate, as the device perfectly simulates its own noise, while the average code fidelity estimation for assessing performance is efficient by using randomized benchmarking-like techniques. Finally, unlike other approaches to error correction, QVECTOR does not assume any error model \emph{a priori}. In contrast, it tailors the encoding and recovery processes to the noise inherent in the device, which might allow to correct errors beyond the Pauli-error model used by stabilizer codes. 


We simulated two distinct five-qubit examples of the QVECTOR algorithm which can be performed with existing hardware. In the first case, we simulated an encoding of a single qubit into three physical qubits subject to realistic rates of phase damping noise. Two ancillary qubits are used to implement the encoded recovery. The simulated QVECTOR algorithm learns an encoding and recovery process which extends the $T_2$ of the quantum memory from $\sim 19~\mu s$ to $\sim 110~\mu s$. Additionally, we considered a quantum communication setting, in which there is no active error correction, but, instead, the quantum information is recovered after a known wait time. Here, we simulated five qubits subject to a combination of continuous amplitude- plus phase-damping noise for various durations. We found that the QVECTOR-learned encoding-decoding pairs continue to bear an advantage beyond wait times of $3.5~\mu s$, the point where the five qubit code fails to be useful. By testing the QVECTOR-learned encoding-decoding pairs under a Pauli-twirled approximation of the same error model, we found that QVECTOR may outperform standard stabilizer codes by exploiting coherence among Pauli errors.

Although promising, the simulation results neither prove the algorithm's scalability nor render it impractical. Reaching such conclusions will likely require an empirical approach. The outcome will be largely dependent on the nature of the cost function landscape (i.e. the estimated average code fidelity as a function of the circuit parameters) and the classical optimization methods that are used to explore it. It is possible that, as the system size is scaled up, the cost function landscape becomes increasingly proliferated with poor local optima or saddle points.

Conversely, and more optimistically, it is possible that realistic noise processes will possess more structure to exploit than the simulated noise models. This could result in a cost function landscape with many favorable local optima. An inspiring example is given in recent work where the structure of correlated noise is exploited to design non-standard quantum error correction schemes for quantum sensing \cite{Layden2017}. 
Crucially, the nature of this cost function landscape is highly dependent on the choice of variational circuit structure, highlighting the importance of designing effective variational circuits.


There are several issues which remain to be investigated. First, the algorithm should be simulated against a more realistic noise model which takes into account the error of each gate, including those of state preparation and measurement. 
There will be 
stochastic errors in the cost function estimation not yet accounted for due to finite sampling error and state preparation and measurement (SPAM) errors. For the latter, it may be favorable to use randomized benchmarking techniques \cite{Magesan2012}, fitting a function of the fidelity estimation for various circuit iterations to obtain a more accurate error rate per iteration estimate, mitigating the offset due to SPAM errors.

There are several other metrics for scoring the performance of a quantum error correction scheme in addition to the simple-to-compute average fidelity. One figure of merit, known as the worst-case fidelity, or simply the error rate \cite{Sanders2015}, is considered to be a more faithful metric for the quality of a quantum process \cite{Wallman2014}, although much more difficult to calculate. However, recent work has shown that worst-case fidelity, along with average fidelity fail to properly assess the performance of error correcting schemes in some cases \cite{Willsch2017}. It remains to determine which metrics will ultimately be the most useful in practice. 


The most important future direction is the augmentation of the existing QVECTOR algorithm to enable learning error-corrected quantum gates for fault-tolerant quantum computation. Towards this, we may view the current work as a step in this direction, as it provides evidence that simple error-corrected gates (i.e. the identity gate) may be variationally constructed. One could imagine training the encoded recovery circuit to be able to apply a target elementary logical gate, such as a two-qubit CNOT gate. In principle, a polynomially-sized universal gate set could be learned in this manner, and then implemented for quantum computation. The efficacy of this approach will be determined by the performance of such gates under composition.

Concatenation of quantum error correcting codes provides the basis for the standard model of quantum computation. It remains to explore concatenated variational error correcting codes. As an example, one could imagine learning several five-qubit variational codes and corresponding recovery circuits, placing five of these in parallel, and applying another round of variational encoding on these twenty-five qubits.





Finally, the most encouraging feature of this approach to quantum error correction is that it can be implemented with existing hardware. Although it lacks the beauty of stabilizer QEC and does not boast any theoretical promises \`{a} la threshold theorems, QVECTOR is sufficiently different from standard approaches that it may provide a fresh avenue of exploration towards realizing quantum computation.

\vspace{.5cm}
\noindent \textbf{Acknowledgements} We acknowledge useful discussions with Tim Menke, David Layden, Morten Kj\ae rgaard, Amara Katabarwa, Sukin Sim, Robin Blume-Kohout, Nicolas Sawaya, Kevin Obenland, and Lorenza Viola. We are especially thankful to Pierre-Luc Dallaire-Demers for pointing us in the right direction for estimating fidelity. AA-G, PDJ, and JO acknowledge support from the Vannevar Bush Faculty Fellowship program sponsored by the Basic Research Office of the Assistant Secretary of Defense for Research and Engineering and funded by the Office of Naval Research through grant N00014-16-1-2008. JR and AAG acknowledge the Air Force Office of Scientific Research for support under Award: FA9550-12-1-0046. YC and AAG acknowledge NSF Quantum Information and Quantum Computation for Chemistry: Challenges and Opportunities,  Grant \# CHE-1655187.  JO acknowledges support from Lincoln Laboratory, contract number 7000381754.
AA-G acknowledges the Army Research Office under Award: W911NF-15-1-0256.

\vspace{1cm}

\bibliographystyle{unsrt}
\bibliography{main.bbl}

\end{multicols}

\newpage
\appendix

\section*{Appendix}

\section{Noise models}
\label{app:noisesimulation}

We simulated the action of noise channels using the standard Kraus operator formalism. In the one qubit case, the action of the channel in the density matrix is given by the operation 
\begin{align}
\mathcal{N}(\rho) = \sum^{m}_{j=1} K_j \rho K_j^{\dagger}
\end{align}
where $\rho_i$ is the initial density matrix and $K_j$ are the corresponding Kraus operators satisfying the completeness relation $\sum^{m}_{j=1} K_j^{\dagger} K_j = I$. To simulate the effect of noise on an $n$-qubit register, we assumed an independent noise model, where the Kraus operators correspond to a tensor products of single qubit Kraus operators. 
The effect of noise on the $n$-qubit register can be written as:
\begin{align}
\mathcal{N}^{\otimes n}(\rho) = \sum_{j_1, \ldots, j_N} \left(\bigotimes_{i=1}^n K_{j_i}\right) \rho \left(\bigotimes_{i=1}^n K_{j_i}^{\dagger}\right),
\end{align}
where the sum runs over all the possible $m^N$ tuples of $j_1,\cdots,j_N$ indexes. 

We consider two different noise channels in our QVECTOR simulations: phase-damping (PD) and a combination of amplitude damping and phase damping (APD). Both are captured by standard $T_{1,2}$ decoherence according to the following map:
\begin{align}
& \rho = \begin{bmatrix} 1-\rho_{11} & \rho_{01} \\ \rho_{01}^{*} & \rho_{11} \end{bmatrix} \rightarrow \\ 
& \begin{bmatrix} 1-\rho_{11}e^{-t/T_1} & \rho_{01}e^{-t/2T_1}e^{-t/T_{\phi}} \\ 
\rho_{01}^{*}e^{-t/2T_1}e^{-(t/T_{\phi})} & \rho_{11} e^{-t/T_1}\end{bmatrix}
\end{align}
where $\frac{1}{T_{\phi}} = \frac{1}{T_2} - \frac{1}{2T_1}$. The Kraus operators for each channel are described in Table \ref{KrausOps}. 
The parameters $\gamma$ and $\lambda$ 
are computed with respect to the experimental parameters corresponding to the wait time of the decoherence, ($t_{\textup{step}}$), the decay time, ($T_1$), and the dephasing time, ($T_2$), according to the equations $\gamma=1-e^{-t_{\textup{step}}/T_1}$ and $\lambda=e^{-t_{\textup{step}}/T_1}+e^{-2t_{\textup{step}}/T_2}$. We also employed the Pauli-twirled version of the APD channel, denoted as PTA-APD, as described in \cite{Geller2013}. The PD channel is obtained by assuming $t_{\textup{step}},T_2\ll T_1$ and the phase-error probability per time step is $p=(1-e^{-t_{\textup{step}}/T_2})/2$. This assumption does not fully capture the specifications of the Xmon qubits which we are basing the simulations on \cite{Barends2013,Barends2014}. This model, therefore, is chosen as a simple starting point from which more-sophisticated explorations should be carried out. 

In our simulations, we chose noise specifications to match error rates of the physical processes arising in state of the art superconducting qubit architectures. In the first simulation, we modeled the noise as independent single-qubit PD channels, acting on the three code-carrying qubits prior to the recovery circuit. The single-qubit phase-error probability per recovery step was set so as to incorporate the error incurred by each gate in the recovery circuit
as follows. 

The three-qubit noise process $\mathcal{N}$ was constructed from a circuit of single qubit phase-flip processes, which correspond to the unitary gates in the recovery circuit. To each single-qubit gate in the recovery circuit, we associated a corresponding single-qubit phase-flip noise process. The phase-flip probability of this noise process was set to $p'=0.00110$ to match the single-qubit gate error rate reported in \cite{Barends2014}. To each two-qubit gate in the recovery circuit, we associated a corresponding pair of parallel single-qubit phase-flip noise processes. This phase flip probability was set to $p''=0.00113$, as computed from the two-qubit gate error rate of \cite{Barends2014}. 

Taking the product of these sequences of phase-flip noise processes, we compute that, throughout the recovery circuit, each qubit incurs a probabilistic phase-error rate of $p=0.091$. The global noise process $\mathcal{N}$, then, consists of three independent PD channels, each with error probability $p=0.091$.

In the case where a single physical qubit is used as a quantum memory without error correction, the error rate will be smaller, as the noisy gates are not present. The error model in this case is single-qubit phase damping with $T_2=19~\mu s$. The performance of the no-encoding case is fairly compared to the error correction case by choosing the duration of this phase-damping decay to match the duration of the recovery circuit. Taking the single-qubit and two-qubit gate times to be $10~ns$ and $40~ns$, respectively, as reported in \cite{Barends2014}, and taking the specifications of the variational recovery circuit described in Appendix \ref{app:simulation}, the recovery duration is computed to be $1.8~\mu s$. Accordingly, the no-encoding physical qubit noise process over this time step is probabilistic phase damping with error probability $p=0.045$.

In the second simulation, we considered a setting where no active recovery is available and APD noise acts continuously during the wait time between encoding and decoding. We chose the value $T_2=19~\mu s$ to match the dephasing rates reported in \cite{Barends2014}. Although the $T_1$ times in this work were roughly twice that of $T_2$, we consider a model in which $T_1=57~\mu s$, three times that of $T_2$. We choose this regime so that the error model differs sufficiently from isotropic depolarizing noise, in which the five-qubit code is known to be optimal \cite{Reimpell2005}. In this regime, the dephasing errors are more dominant than the amplitude damping errors (which are coherent combinations of Pauli $X$ and $Y$ errors).

\begin{table*}[!]
\centering
\begin{tabular}{|c|c|}
\hline
\textbf{Noise channel} & \textbf{Kraus Operators} \\
\hline
Phase damping (DP) & $K_1 = \sqrt{p} I; \quad K_2 = \sqrt{1-p} Z$ \\
\hline
Amplitude- plus Phase-Damping (APD) & $K_1=
  \begin{bmatrix} 1 & 0 \\
   0 & \sqrt{1-\gamma-\lambda} \\
  \end{bmatrix}; K_2=\begin{bmatrix}
   0 & \sqrt{\gamma} \\
   0 & 0 \\
  \end{bmatrix}; K_3 =\begin{bmatrix}
   0 & 0 \\
   0 & \sqrt{\lambda} \\
  \end{bmatrix}$ \\
\hline
Pauli twirled - APD (PTA-APD) &  $K_1 = (1-p_X-p_Y-p_Z)I$; \quad $K_2=p_X X$; \quad $K_3=p_Y Y$; \quad $K_4=p_Z Z$\\
\hline
\end{tabular}
\caption{Kraus Operators for common one-qubit error channels employed in classical simulations of the QVECTOR protocol. $p$ is the error rate for the phase damping process. The parameters $\gamma$ and $\lambda$ are associated to the amplitude damping and phase damping processes, respectively, and are computed from the process time $t_{\textup{step}}$ and the T$_1$ and T$_2$ times as $\gamma=1-e^{-t_{\textup{step}}/T_1}$ and $\lambda=e^{-t_{\textup{step}}/T_1}+e^{-2t_{\textup{step}}/T_2}$. The parameters for the Pauli-twirled approximation (PTA) of APD are calculated as $p_X=p_Y=\frac{\gamma}{4}$ and $p_Z=\frac{1}{2}-\frac{\gamma}{4}-\frac{\sqrt{1-\gamma-\lambda}}{2}$ \cite{Geller2013}.}\label{KrausOps}
\end{table*}

\section{Numerical simulation of QVECTOR}
\label{app:simulation}

We simulated the QVECTOR protocol using a Python script supplemented with the QuTiP library \cite{johansson2012,johansson2013}. Since our simulations involve a single logical qubit, average fidelities were computed over the one qubit stabilizer states $|s_i\rangle$: $|+\rangle$, $|-\rangle$, $|0\rangle$, $|1\rangle$, $|+i\rangle$ and $|-i\rangle$, which form a 2-design. For systems with 2 or more logical qubits, we could use an approximate 2-design circuit such as the one described in \cite{Nakata2017} (See Appendix \ref{app:fidelityest}).

The simulation of the QVECTOR protocol comprises: 1) preparation of the initial state for the code register $\rho_{s_i}=|s_i\rangle\langle s_i|\otimes| 0^{(n-k)}\rangle\langle 0^{(n-k)}|$, 2) application of the encoding channel $\mathcal{V}_{\vec{p}}$, 3) application of the noise process $\mathcal{N}$, 4) application of the recovery channel $\mathcal{W}_{\vec{q}}$ with $r$ refresh qubits and 5) decoding by application of $\mathcal{V}_{\vec{p}}^{\dagger}$. The final average fidelity is computed as:
\begin{align}\label{fidelitySimulations}
\langle F\rangle = \frac{1}{N_s} \sum^{N_s}_{i=1} \text{F}\left(\rho_{s_i}, \mathcal{V}_{\vec{p}}^{\dagger}\trb{R}{\mathcal{W}_{\vec{q}} \left( \mathcal{N}\mathcal{V}_{\vec{p}} \left(\rho_{s_i} \right) \otimes \ket{0^{(r)}}\bra{0^{(r)}} \right)} \right)
\end{align}
where $N_s$ stands for the number of states employed to compute the average fidelity and $\text{F}(\rho,\sigma) = \textup{tr}(\sqrt{\sigma^{1/2}\rho\sigma^{1/2}})^2$. The partial trace is performed over the refresh qubits used for recovery, which are initialized in the state $\ket{0^{(r)}}$. In the simulations presented in this work $N_s=6$, corresponding to state 2-design formed by the six one-qubit stabilizer states. For our example of the phase damping channel, $n=3$, $k=2$ and $r=2$. In our example of APD, we explored an approach without recovery, such that $n=5$, $k=4$, and $r=0$.

\begin{figure*}[ht]
\begin{center}
\includegraphics[width=16cm]{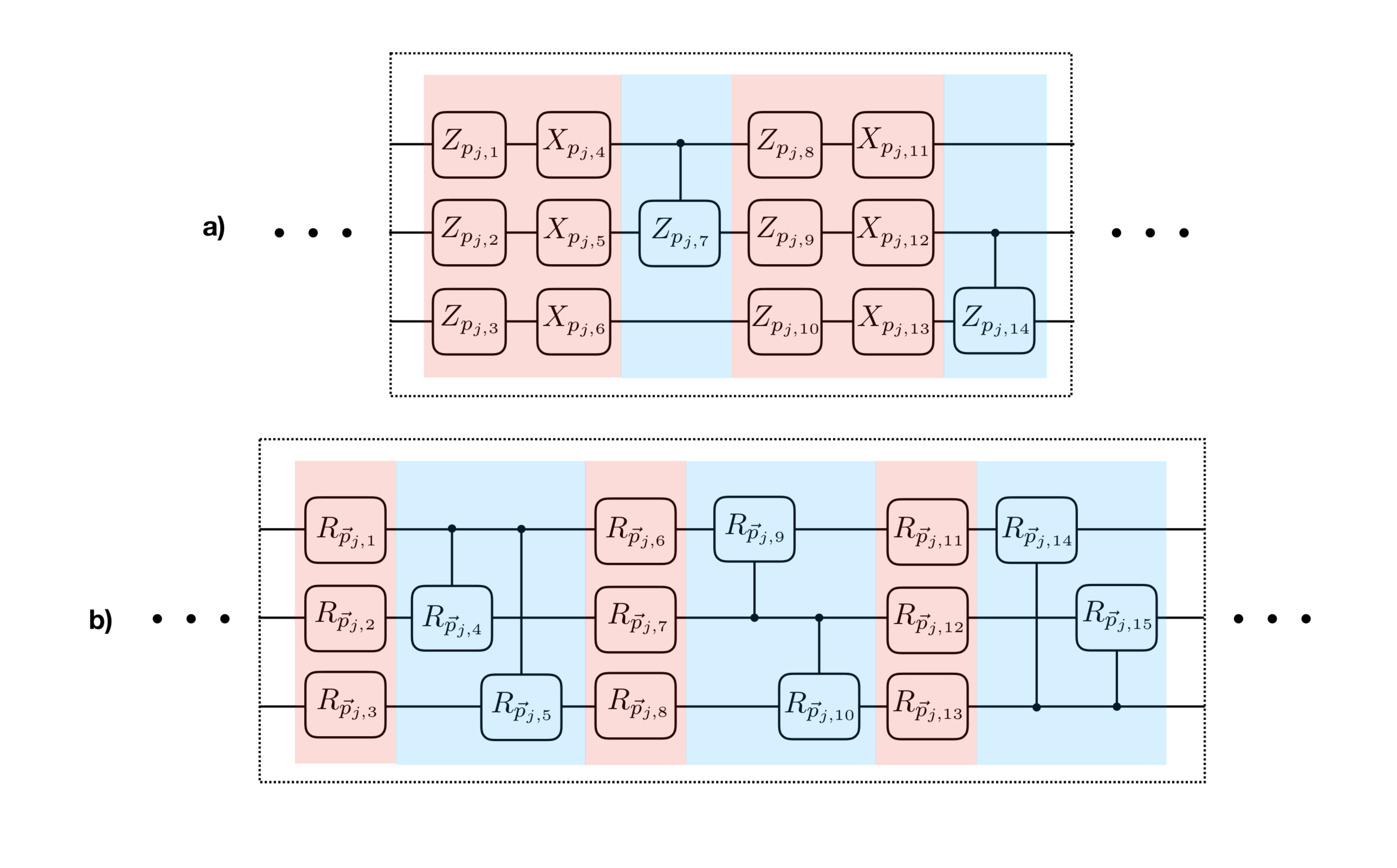}
\end{center}
\caption{Unit-cells of programmable circuits employed for QVECTOR simulations for a three-qubit register. Both circuits comprise alternating layers of single-qubit rotations (red) and entangling operations (blue). a) Unit-cell of adjacent controlled-$Z_{\theta}$ rotations interleaved with layers of $Z_{\theta}$ and $X_{\theta}$ rotations, where $Z_{\theta} = \exp(-i\frac{\theta}{2}Z)$ (Analogous for X). b) Unit-cell comprises all the possible controlled-arbitrary single-qubit rotations (denoted by $R_{\vec{p}}$) in a given register, interleaved with layers of arbitrary single qubit rotations. Here $R_{\vec{p}}=Z_{p_1}X_{p_2}Z_{p_3}$. The index $j$ runs from 1 to $l$, where $l$ is the number of repetitions of the unit-cell. All the circuits are complemented with a single-qubit rotation layer after the last repetition.}\label{circuitHeuristics}
\end{figure*}

The encoding $\mathcal{V}_{\vec{p}}$ and recovery $\mathcal{W}_{\vec{q}}$ were implemented using programmable circuits  \cite{Sousa2007,Daskin2012}. These types of circuits generally comprise a fixed networks of gates, where the parameters associated to the gates, i.e. rotation angles, constitute the variables for optimization. The pattern defining the network of gates is regarded as the unit-cell, which can be repeated to increase the flexibility of the model. The programmable circuits employed in this work are illustrated in Figure \ref{circuitHeuristics}. The unit-cells of these circuits follow a pattern consisting of layers of single qubit rotations interleaved with entangling blocks. This heuristic construction is very amenable to current quantum hardware \cite{Kandala17, Moll2017}.

Our first programmable circuits, shown in Figure \ref{circuitHeuristics}(a), are comprised of two layers of single-qubit rotations and two entangling layers. The layers of single-qubit rotations contain $X_{\theta}$ and $Z_\theta$ rotations applied on each qubit, where the notation $A_{\theta}$ stands for $\exp(-i\frac{\theta}{2}A)$. The entangling layers comprise all the possible adjacent controlled-$Z_{\theta}$ gates, with controls in the odd (even) qubits, and targets in the even (odd) qubits. The total number of parameters in this circuit is $2n+l(5n-1)$ where $l$ is the number of repetitions of the unit cell. The unit-cell of our second programmable circuit, shown in Figure \ref{circuitHeuristics}(b), comprises layers of single qubit arbitrary rotations interleaved with controlled- single qubit arbitrary rotations from the $i$-th qubit to the rest of the qubits in the register. Arbitrary single-qubit rotations were implemented as $R_{\vec{p}}=Z_{p_1}X_{p_2}Z_{p_3}$. The total number of parameters of this circuit is $3ln(2n-1)+3n$. In the simulations of the phase damping channel, we represented the circuit for $\mathcal{V}_{\vec{p}}$ an $\mathcal{W}_{\vec{q}}$ using 10 and 15 repetitions of the circuit of Figure \ref{circuitHeuristics}(a). For the encoder of the APD example, we employed three repetitions of the unit-cell of Figure \ref{circuitHeuristics}(b).

After determining the form of the unitaries for $\mathcal{V}_{\vec{p}}$ and $\mathcal{W}_{\vec{q}}$, the QVECTOR simulation proceeds by optimizing the fidelity in Eq. \ref{fidelitySimulations}. For our numerical simulations, we employed the L-BFGS method \cite{byrd95} with a numerical gradient (central finite difference formula with step size $h=10^{-6}$). The circuit parameters were initialized by generating 100 random parameter settings, drawn uniformly from the range $[0,4\pi)$, and selecting the set with the highest fidelity. Several optimizations were launched in parallel and the best result was selected. Our numerical explorations indicated that the average fidelity cost function might contain several local optima, and sampling different initial points for the optimization might benefit the success of the procedure. 

We point out that in experimental implementations of QVECTOR, the fidelity cost function will be affected by errors introduced by sampling, as well as SPAM errors. In this scenario, the procedure might benefit from the use of optimization algorithms more tolerant to noise, such as  Simultaneous Perturbation Stochastic Approximation (SPSA) \cite{Moll2017}, as well as from global optimization techniques such as Basin-Hopping \cite{Wales1997}. Additionally, we expect that, in order to more-effectively estimate the average fidelity in the presence of these unwanted fluctuations, randomized benchmarking techniques should be employed (as outlined in the outlook section).

Lastly, we describe the bi-convex optimization routine which was used to compute the optimal average fidelities for the APD noise processes at different wait times. This method developed in \cite{Kosut2009} takes advantage of the fact that the average fidelity metric is a bi-linear, and, therefore, bi-convex function of the encoding and decoding channels,
\begin{align}
\langle F(p\mathcal{D}+(1-p)\mathcal{D}',\mathcal{N},q\mathcal{E}+(1-q)\mathcal{E}')\rangle
\geq p\langle F(\mathcal{D},\mathcal{N},q\mathcal{E}+(1-q)\mathcal{E}')\rangle
+(1-p)\langle F(\mathcal{D}',\mathcal{N},q\mathcal{E}+(1-q)\mathcal{E}')\rangle\nonumber\\
\langle F(p\mathcal{D}+(1-p)\mathcal{D}',\mathcal{N},q\mathcal{E}+(1-q)\mathcal{E}')\rangle
\geq q\langle F(p\mathcal{D}+(1-p)\mathcal{D}',\mathcal{N},\mathcal{E})\rangle
+(1-q)\langle F(p\mathcal{D}+(1-p)\mathcal{D}',\mathcal{N},\mathcal{E}')\rangle,\nonumber\\
\end{align}
where, although these hold with equalities, the inequalities  suffice to enable the convex optimization method. The method proceeds by first choosing a random initial encoding $\mathcal{E}$ (chosen as an isometry), and then performing semidefinite programming to optimize the average fidelity with respect to the decoding $\mathcal{D}$, which is implemented using CVX, a package for specifying and solving convex programs \cite{Grant2008a,Grant2008b}. Then, setting the decoding to this optimized variable, the average fidelity is convex-optimized with respect to the encoding map $\mathcal{E}$. This process is iterated, with average fidelity increasing in each step until the improvement in a step falls below a chosen threshold value. In practice, the optimized average fidelity varies from one run to the next depending on the initial encodings (the procedure is, otherwise, deterministic). To improve confidence that the obtained vale is sufficiently close to the optimal value, we perform many runs, decreasing the threshold value until the obtained average fidelities become sufficiently independent of the choice of threshold. Accordingly, although we can build substantial evidence for the value being close to optimal, this method can only definitively obtain a lower bound on the optimal average fidelity.

\section{Accuracy of average fidelity estimate from approximate unitary 2-design}
\label{app:fidelityest}

\begin{figure*}[ht]
\begin{center}
\includegraphics[width=12cm]{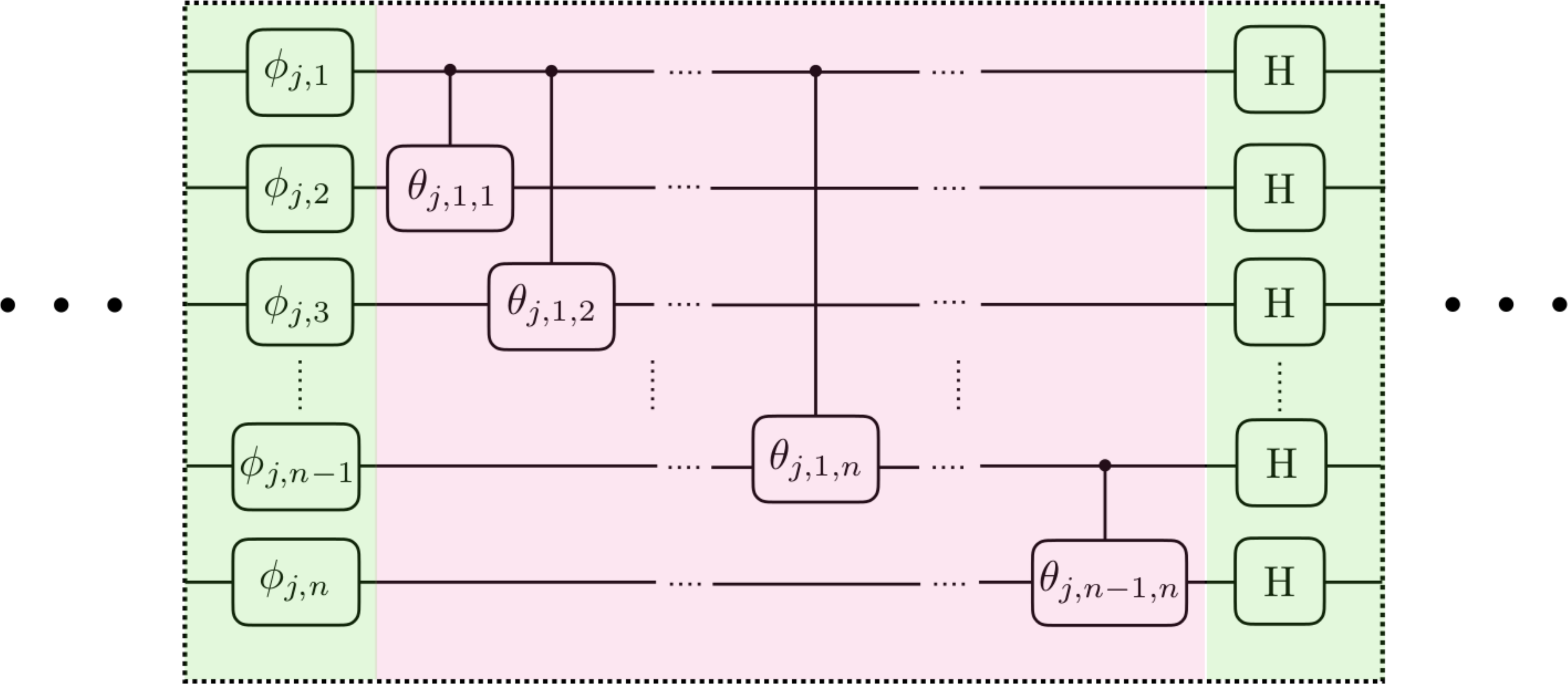}
\end{center}
\caption{Unit-cell of the quantum circuit that implements an approximate unitary 2-design according to \cite{Nakata2017} in an $n$-qubit register. Layers are separated by colors. The one- and two-qubit gates in the first two layers correspond to diag(1, $e^{i\phi_{j,p}}$) and diag(1, 1, 1, $e^{i\theta_{j,p,q}}$), respectively. The phases $\phi_{j,p}$ and $\theta_{j,p,q}$ are chosen from $\{0, 2\pi/3, 4\pi/3\}$ and $\{0, \pi \}$, respectively, uniformly at random. The third layer comprises Hadamard gates ($\text{H}$) applied to all the qubits. All the gates in the first and the second layers are diagonal in the Pauli-Z basis and can be applied simultaneously. The index $j$ runs from 1 to $\ell$, where $\ell$ is the number of repetitions of the unit-cell. 
}\label{nakataCircuit}
\end{figure*}

First we prove that the approximate 2-design of \cite{Nakata2017} (See Figure \ref{nakataCircuit}), with $\ell$ applications of the randomization circuit, leads to an estimator of the true average fidelity with bias upper bounded by $\frac{2^{k(\ell+1)}+2^{k\ell}-2}{2^{2k\ell}(2^{k}-1)}\sim \mathcal{O}(1/2^{k\ell})$. Let $\nu_{\ell}$ be the measure on the unitary group that is sampled from using the approximate 2-design of \cite{Nakata2017} with $\ell$ repetitions. In the average fidelity estimation scheme, if we draw from $\nu_{\ell}$ instead of an exact 2-design (such as the Haar measure), we will be sampling from a biased estimator with mean
\begin{equation}
\label{eq:biasedestimator}
\langle E_{\ell}\rangle = \int\bra{0}U^{\dagger}\mathcal{M}\left(U\ketbra{0}U^{\dagger}\right)U\ket{0}d\nu_{\ell}(U),
\end{equation}
where $\mathcal{M}=\mathcal{V}_{\vec{p}}^{\dagger}\mathcal{W}_{\vec{q}}\mathcal{V}_{\vec{p}}$. We give an upper bound on the bias $|\langle F\rangle -\langle E_{\ell}\rangle|$. The 2-design average over this measure may be renormalized in order to interpret it as a quantum channel $\mathcal{R}_{\ell}(\sigma)=d\int U^{\otimes 2} \sigma {U^{\otimes 2}}^{\dagger}d\nu_{\ell}(U)$.
As shown in \cite{Nakata2017}, this quantum channel can be written as a convex combination of two quantum channels, 
\begin{equation}
\label{eq:Nakatachannel}
\mathcal{R}_{\ell}(\sigma) = (1-p_{\ell})\mathcal{G}(\sigma)+p_{\ell}\mathcal{C}_{\ell}(\sigma),
\end{equation}
where $\mathcal{G}$ is the renormalized average of an exact 2-design, $\mathcal{C}_{\ell}$ is a quantum channel, and $p_{\ell}=\frac{d^{\ell+1}+d^{\ell}-2}{d^{2\ell}(d-1)}$, with $d=2^{k}$ being the Hilbert space dimension. To leverage this quantum channel interpretation of the 2-design, we rewrite the expression in Equation \ref{eq:biasedestimator} as follows,
\begin{align}
\langle E_{\ell}\rangle &= \int\trb{}{\mathcal{M}\otimes\mathcal{I}\left(U\otimes U\ketbra{00}U^{\dagger}\otimes U^{\dagger}\right)\mathbb{F}}d\nu_{\ell}(U)\\
&=\frac{1}{d}\trb{}{\mathcal{M}\otimes\mathcal{I}\left(\mathcal{R}_{\ell}(\ketbra{00})\right)\mathbb{F}},
\end{align}
where $\mathbb{F}$ is the swap operator on the two systems. Replacing the channel $\mathcal{R}_{\ell}$ with the convex combination in Equation \ref{eq:Nakatachannel}, we obtain an expression for the estimator mean in terms of the actual mean,
\begin{align}
\langle E_{\ell}\rangle &=\frac{1}{d}(1-p_{\ell})\trb{}{\mathcal{M}\otimes\mathcal{I}\left(\mathcal{G}(\ketbra{00})\right)\mathbb{F}}\\
&+\frac{1}{d}p_{\ell}\trb{}{\mathcal{M}\otimes\mathcal{I}\left(\mathcal{C}_{\ell}(\ketbra{00})\right)\mathbb{F}}\\
&=(1-p_{\ell})\langle F\rangle +p_{\ell}\delta_{\ell}.
\end{align}
The bias of the estimator is $|\langle F\rangle -\langle E_{\ell}\rangle|=p_{\ell}|\langle F\rangle-\delta_{\ell}|$. To bound this value, we bound
\begin{align}
\delta_{\ell}=\frac{1}{d}\trb{}{\mathcal{M}\otimes\mathcal{I}\left(\mathcal{C}_{\ell}(\ketbra{00})\right)\mathbb{F}}.
\end{align}
From Equation (12) in \cite{Nakata2017}, 
\begin{align}
\mathcal{C}_{\ell}(\ketbra{00})=\alpha P_{\textup{sym}}+\beta\sum_i\ketbra{ii},
\end{align}
where $\alpha,\beta\geq 0$ and $P_{\textup{sym}}$ is the projector into the symmetric subspace. Since $\mathcal{C}_{\ell}$ is separable, it is invariant under partial transpose of either system. The partial transpose of the swap operator is the unnormalized Bell state $\mathbb{F}^{T_B}=\sum_{i,j}\ket{ii}\bra{jj}\equiv d\Omega$. Since the trace of a bipartite operator is equal to the trace of the partial transpose of that bipartite operator, we can use $\mathcal{C}_{\ell}(\ketbra{00})^{T_B}=\mathcal{C}_{\ell}(\ketbra{00})$ and $\frac{1}{d}\mathbb{F}^{T_B}= \Omega$ to obtain
\begin{align}
\delta_{\ell}=\trb{}{\mathcal{M}\otimes\mathcal{I}\left(\mathcal{C}_{\ell}(\ketbra{00})\right)\Omega}.
\end{align}
Observing that $\trb{}{\mathcal{M}\otimes\mathcal{I}\left(\mathcal{C}_{\ell}(\ketbra{00})\right)\Omega}$ is the inner product of two quantum states, we can upper bound this value by 1. Thus, the bias of the estimator is upper bounded by $p_{\ell}=\frac{d^{\ell+1}+d^{\ell}-2}{d^{2\ell}(d-1)}\sim \mathcal{O}(1/d^{\ell})$.

After $N$ samples from this biased estimator, our estimated average fidelity value is expected to deviate from the estimator mean $\langle E_{\ell}\rangle$ by $\sqrt{\langle E_{\ell}\rangle(1-\langle E_{\ell}\rangle)/N}$. An upper bound on the expected deviation of the sampling-estimated  average fidelity from the true average fidelity is 
\begin{equation}
\frac{1}{\sqrt{N}}+\frac{d^{\ell+1}+d^{\ell}-2}{d^{2\ell}(d-1)}\sim \mathcal{O}\left(\frac{1}{\sqrt{N}}+\frac{1}{d^{\ell}}\right).
\end{equation}

\clearpage

\end{document}